\documentclass[11pt]{revtex4}

\usepackage{graphicx}

\usepackage{hyperref}
\usepackage{amsmath,amsfonts,amssymb}
\hypersetup{dvips,dvipdfm,colorlinks=true,urlcolor=magenta,filecolor=magenta,linktoc=page,citecolor=red,linkcolor=blue,bookmarks=true}
\usepackage{graphicx,epstopdf}
\usepackage{array}
\newcounter{defcounter}
\newcolumntype{L}[1]{>{\raggedright\let\newline\\\arraybackslash\hspace{0pt}}m{#1}}
\newcolumntype{C}[1]{>{\centering\let\newline\\\arraybackslash\hspace{0pt}}m{#1}}
\newcolumntype{R}[1]{>{\raggedleft\let\newline\\\arraybackslash\hspace{0pt}}m{#1}}

\begin{document}
	
	\title{ Thermodynamics of the dissipative cosmic fluid and stability criteria\\
		}
	\author{Pritikana Bhandari\footnote {pritikanab@gmail.com}}
	\affiliation{Department of Mathematics, Jadavpur University, Kolkata-700032, West Bengal, India.}
	
	\author{Sourav Haldar\footnote {sourav.math.ju@gmail.com}}
	\affiliation{Department of Mathematics, Jadavpur University, Kolkata-700032, West Bengal, India.}
	
	\author{Subenoy Chakraborty\footnote {schakraborty.math@gmail.com}}
	\affiliation{Department of Mathematics, Jadavpur University, Kolkata-700032, West Bengal, India.}

	\begin{abstract}
		
		The work deals with the thermodynamical aspects of the cosmic substratum which is dissipative in nature. For homogeneous and isotropic model of the Universe this dissipative phenomenon is effective bulk viscous pressure in nature and is related to the particle creation mechanism. Finally, the stability criteria for the thermal equilibrium has been analyzed and are presented in tabular form restricting the particle creation parameter or the variation of the equation of state parameter with the volume. \\\\
		{\bf Keywords\,:} Thermodynamical equilibrium, Dissipative cosmic fluid, Equation of State, Particle creation mechanism.\\\\
		PACS Numbers\,: 95.30.Sf, 98.80.Cq, 11.90.+t, 05.70.Ce\\\\
	
	\end{abstract}

	\maketitle
	
	
	
	
\section{Introduction}

Non-equilibrium thermodynamics is supposed to play a great role in the evolution of the Universe at very early phase. Long back in 1939 such a programme was initiated by Schr\"{o}dinger \cite{Schrodinger1} considering the microscopic description of the gravitationally induced particle production in an expanding Universe. Formally, considering quantum field theory in a curved space time this issue was analyzed by Parker and others \cite{Parker:1969au,Parker:1968mv, Birrell:1982ix, Mukhanov:2007zz}. Subsequently, this thermodynamical issue has been studied considering first and second order theories separately. The first order theory was initiated by Eckart \cite{Eckart:1940te} and Landau and Lipschitz \cite{Landau1} but it has bad features namely causality violation and stability problem. However, these problems were eliminated when second order deviations from equilibrium were considered by Muller \cite{Muller:1967zza}, Israel \cite{Israel:1976tn}, Israel and Stewart \cite{Israel:1979wp,Israel3} and Pavon $et~al$ \cite{Pavon1}. According to them, dissipative phenomena like bulk and shear viscous pressure and heat flux should be considered as dynamical variables having causal evolution equations and the thermal and viscous perturbations propagate at subluminal speeds.

From cosmological stand point, the Universe is usually assumed to be homogeneous and isotropic and consequently, bulk viscous pressure is the sole dissipative phenomenon which may appear either due to coupling of different matter components \cite{Weinberg:1971mx,Straumann:1986zr, Schweizer1, Udey1, Zimdahl:1996fj} or due to non-conservation of (quantum) particle number \cite{Zeldovich:1970si, Murphy:1973zz,Hu:1982pv}. In the present work, second alternative for dissipative effect is considered and for simplicity only adiabatic particle production \cite{Prigogine:1989zz, Calvao:1991wg} is taken into account. Also the entropy flow due to Israel-Stewart \cite{Israel:1979wp} is related to the cosmological (isentropic) particle production. Due to isentropic nature of the thermodynamical system the entropy per particle is constant but entropy production is caused by the enlargement of the phase space of the system $i.e.$ expansion of the Universe. An explicit thermodynamical analysis of the cosmic substratum with dissipative nature due to particle creation is presented and finally thermodynamical stability is shown for different situations. The paper is organized as follows\,: sect. 2 deals with basic equations for the particle creation mechanism. The thermodynamics of the cosmic dissipative fluid has been discussed in sect. 3. Finally, stability criteria has been examined and concluding remarks are presented in sect. 4.

\section{Basic equations for the particle creation mechanism}

In the background of homogeneous and isotropic spatially flat FLRW model of the Universe, the Einstein field equations can be written as
\begin{equation} \label{eq1}
3H^2= \kappa \rho ~~~~\mbox{and}~~~~ 2\dot{H}=-\kappa (\rho + p+ \Pi)
\end{equation}
where ~$\kappa = 8 \Pi G$~ is the Einstein's gravitational constant, $H = \frac{\dot{a}}{a}$ is the Hubble parameter, $a(t)$ is the scale factor of the FLRW model and the cosmic fluid is characterized by the energy-momentum tensor\,:
\begin{equation} \label{eq2}
T_\mu ^\nu = (\rho+p+\Pi) u_\mu u^\nu + (p+\Pi) g_\mu ^\nu \,.
\end{equation} 
Here $\rho$ and $p$ are the usual energy density and thermodynamic pressure of the cosmic fluid, the dissipative term $\Pi$ is the bulk viscous pressure and $u_\mu$ is the fluid particle four velocity vector. So the energy conservation relation ~$T^\nu_{\mu;\nu}=0$~ takes the form 
\begin{equation} \label{eq3}
\dot{\rho}+3H(\rho+p+\Pi)=0\,,
\end{equation}
where an overdot stands for differentiation with respect to the cosmic time `$t$'.

If the thermodynamical system is considered to be open so that fluid particles are not conserved $(N^\mu_{\,;\mu} \neq 0)$ $i.e.$ \cite{Maartens1,Zeldovich:1970si}
\begin{equation} \label{eq4}
\dot{n}+\theta n= n \Gamma
\end{equation}
then non-equilibrium thermodynamics comes into picture and the dissipative term $\Pi$ can be considered as effective bulk viscous pressure due to particle creation. In other words, the cosmic substratum may be chosen as perfect fluid with varying particle number. In the above, `$n$' stands for the particle number density, $\theta = u^\mu _{\,;\mu}$ is the fluid expansion, $\Gamma$ is termed as the rate of change of the particle number in a co-moving volume $V$ and notationally, we write $\dot{n}=n_{,\mu} u^\mu$\,. In the system, there will be particle creation or annihilation depending on the sign of~ $\Gamma$ to be positive or negative respectively.
\vspace{3mm}\\
The first law of thermodynamics 
\begin{equation} \label{eq5}
	dE=dQ-p\,dV\,,
\end{equation}
is essentially the conservation of internal energy $E$ with $dQ$\,, the heat received by the system in  time $dt$\,. Now writing $n=\frac{N}{V}$ and heat per unit particle $dq=\frac{dQ}{N}$ the above conservation equation (\ref{eq5}) becomes the Gibb's equation 
\begin{equation} \label{eq6}
Tds = dq = d\left(\frac{\rho}{n}\right)+p\,d\left(\frac{1}{n}\right) \,.
\end{equation} 
Here `$s$' stands for the entropy per particle, $T$ is the temperature of the fluid, and it is to be noted that this equation also holds for the present open thermodynamical system. Also using the conservation equations (\ref{eq3}) and (\ref{eq4}) one gets the entropy variation from equation (\ref{eq6}) as
\begin{equation} \label{eq7}
nT\dot{s} = -3H\Pi- \Gamma (\rho + p)\,.
\end{equation}
Now assuming the present thermodynamical system to be adiabatic in nature so that the entropy per particle is constant (variable in non-isentropic process) and as a result the effective dissipative pressure is related to the particle creation rate by the relation \cite{Maartens1, Zeldovich:1970si,Chakraborty:2014oya,Chakraborty:2014ora,Chakraborty:2014fia}
\begin{equation} \label{eq8}
\Pi = -\frac{\Gamma}{3H} (\rho + p)\,.
\end{equation}
Thus in an adiabatic process, a perfect fluid with particle creation phenomena is equivalent to a dissipative fluid. Also in this system the entropy production is caused by the enlargement of the phase space (expansion of the Universe in the present model). Further, it is worthy to mention that the equivalent bulk pressure does not correspond to conventional non-equilibrium phase, rather a state having equilibrium properties as well (not the equilibrium era with $\Gamma =0$). Now eliminating the dissipative pressure $\Pi$ from the Einstein field equation (\ref{eq1}) using the isentropic condition (\ref{eq8}) one obtains the evolution equation as \cite{Chakraborty:2014oya,Chakraborty:2014ora,Chakraborty:2014fia}
\begin{equation} \label{eq9}
\frac{2\dot{H}}{3H^2}=(\gamma+1) \left(\frac{\Gamma}{3H}-1\right)
\end{equation}
where $\gamma =\frac{p}{\rho}$ is the equation of state parameter.

\section{Thermodynamics of the cosmic dissipative fluid}

The present section deals with the thermodynamics of the dissipative cosmic fluid (having energy-momentum tensor (\ref{eq2})) in the background of flat FLRW space-time (for thermodynamics of dissipative fluid one may see references \cite{Mathew:2014gpa,Radicella:2014nka,Mazumder:2010zz,Mazumder:2010tq}). The internal energy of the cosmic fluid is described as
\begin{equation} \label{eq10}
E=\rho c^2 V
\end{equation}
with co-moving volume $V=a^3(t)V_0$ (suffix `0' stands for the present value and we choose $a_0 =1$ by convention). Using Clausius relation the first law of thermodynamics ($i.e.$ eq.\,(\ref{eq5})) can be rewritten as
\begin{equation} \label{eq11}
TdS=dQ=dE+\widetilde{p}\,dV\,.
\end{equation}
where~ $\widetilde{p}=p+\Pi$~ is the effective pressure.

If we assume that the Universe experiences a reversible adiabatic expansion then there will be no heat flow ($i.e.$ $dQ = 0$) and as a result the above first law of thermodynamics ($i.e.$ eq.\,(\ref{eq11})) gives the energy-momentum conservation relation (\ref{eq3}). Choosing $T$ and $V$ to be the basic thermodynamical variables $i.e.$ $\rho =\rho (T,V)$ the exact differential criteria for $dS$ in equation (\ref{eq11}) gives \cite{Weinberg:1971mx, Barboza:2015rsa}
\begin{equation} \label{eq12}
d \ln T = d \ln |1+\omega| -\omega \,d \ln V
\end{equation}
where \begin{equation} \label{eq13}
\omega=\frac{\widetilde{p}}{\rho}=\gamma+\frac{\Pi}{\rho}
\end{equation}
is the effective equation of state parameter for the dissipative fluid. Now using the energy conservation equation (\ref{eq3}) one can rewrite equation (\ref{eq12}) as 
\begin{equation} \label{eq14}
d\ln T=d \ln |1+\omega| +d \ln \rho+ d\ln V
\end{equation}
which on integration gives 
\begin{equation} \label{eq15}
(1+\omega)\frac{\rho V}{T}=(1+\omega _0)\frac{\rho _0 V_0}{T_0}=\mbox{constant}
\end{equation}
or equivalently 
\begin{equation} \label{eq16}
E=E_0\left(\frac{1+\omega _0}{1+\omega}\right)\frac{T}{T_0}\,.
\end{equation}
This is the gas law for the dissipative cosmic substratum and it can be named modified ideal gas law.

We now introduce important thermodynamical parameters namely heat capacity parameters, thermal expansivity and compressibility. The fluid's heat capacity at constant pressure is defined as \cite{Callen1}
\begin{equation} \label{eq17}
C_p=\left(\frac{\partial h}{\partial T}\right)_p~,
\end{equation}
where
\begin{equation} \label{eq18}
h=E+\widetilde{p}\,V
\end{equation} 
is the enthalpy of the fluid. Using the above gas law (\ref{eq16}) the explicit form of $C_p$ is
\begin{equation} \label{eq19}
C_p=(1+\omega) \frac{E}{T}=(1+\omega _0) \frac{E_0}{T_0}= \mbox{constant}\,.
\end{equation}
Similarly, using the gas law (\ref{eq16}) and the integrability condition (\ref{eq12}), the fluid's heat capacity at constant volume has the explicit expression
\begin{equation} \label{eq20}
C_v=\left(\frac{\partial E}{\partial T}\right)_V=\frac{d\ln V}{(1+\omega)d\ln V-d\ln \omega}C_p\,.
\end{equation}

On the other hand, choosing $p$ and $T$ as the independent thermodynamic variables the variation of volume can be written as
\begin{equation} \label{eq21}
\frac{dV}{V}=\left(\alpha \,dT-\kappa _T \,dp\right)
\end{equation}
where ~$\alpha =\frac{1}{V}\left(\frac{\partial V}{\partial T}\right)_p$~ is termed as thermal expansivity (at constant pressure) and
\begin{equation} \label{eq22}
\kappa _T = -\frac{1}{V}\left(\frac{\partial V}{\partial p}\right)_T
\end{equation}
is known as isothermal compressibility (at constant temperature). Analogously, one can define the adiabatic compressibility $\kappa _s$ as (keeping entropy instead of temperature to be constant)
\begin{equation} \label{eq23}
\kappa _s=-\frac{1}{V}\left(\frac{\partial V}{\partial{p}}\right)_S\,.
\end{equation}
It should be noted that the ratio of the heat capacities is same as that of the compressibilities $i.e.$ \cite{Kubo1}
\begin{equation} \label{eq24}
\frac{\kappa _s}{\kappa _T}=\frac{C_v}{C_p}
\end{equation}
and is independent of the choice of the cosmic fluid.

From the above definitions of the thermodynamical parameters, it is easy to see that the isothermal compressibility and expansibility are related as \cite{Barboza:2015rsa}
\begin{equation} \label{eq25}
\frac{\alpha}{\kappa _T}=\left(\frac{\partial p}{\partial T}\right)_V\,.
\end{equation}
Now using the modified ideal gas equation (\ref{eq15}) and the integrability condition (\ref{eq12})\,, $\alpha$ can have the explicit expression :
\begin{equation} \label{eq26}
\alpha =\frac{C_p}{(1+\omega) \rho V}\left[1+\frac{d\omega}{\omega \left\{d\omega -\omega (1+\omega) d\ln V \right\}}\right]
\end{equation} 
so that we have 
\begin{equation} \label{eq27}
\kappa _T=\frac{\alpha V}{C_p}
\end{equation}
and
\begin{equation} \label{eq28}
\kappa _s=\frac{\alpha VC_v}{C_P^2}
\end{equation}
where the identity (\ref{eq24}) has been used in deriving equation (\ref{eq28}). Further, using the relation (\ref{eq20}) among the heat capacities, the compressibility coefficients are related by the relation
\begin{equation} \label{eq29}
\kappa _s=\frac{d\ln V}{{(1+\omega) d\ln V-d\ln \omega}}\,\kappa _T\,.
\end{equation}

\section{Stability of the thermodynamical system and concluding remarks}

Normally, in a thermodynamical system work is considered only due to the change in volume. For such system the criteria for stable equilibrium is characterized by the positive definiteness of the second order variation of the internal energy $i.e.$ \cite{Barboza:2015rsa, Kubo1} 
\begin{equation} \label{eq30}
\delta ^2 E=\delta T\,\delta S-\delta p\,\delta V>0\,.
\end{equation}
Now choosing $(T, V)$ or $(S, p)$ as the independent variables for the system the above second order variation can be written as
$$\frac{C_v}{T} \delta T^2+ \frac{1}{V \kappa _T} \delta V^2$$
\vspace{-1.4cm}
\begin{equation} \label{eq31}
\hspace{-3.5cm} \delta ^2 E =~~~~~~~~~~~~\mbox{or} 
\end{equation}
\vspace{-1.2cm}
$$\frac{T}{C_p} \delta S^2+ V \kappa _s \delta p^2 ~.$$
Hence for stability we must have 
\begin{equation} \label{eq32}
C_v, C_p, \kappa_T, \kappa _s \geq 0\,.
\end{equation}
Using these inequalities to the explicit expressions (or relations) for the above thermodynamical parameter for the present dissipative cosmic fluid, the stability criteria for differential range of values of the equation of state parameter `$\omega$' are presented in the following tabular form :\\

\begin{center}
	\begin{table}[ht]
		\renewcommand{\arraystretch}{1.5}
		\caption{Conditions for Stability Criteria} \label{tab:1}
		\begin{tabular}{| >{\centering\arraybackslash}m{5cm}|>{\centering\arraybackslash}m{6cm}|>{\centering\arraybackslash}m{5cm}|}
			\hline
			{\bf Restriction on} $\gamma \,/\,\omega$ & {\bf Constraint on the particle creation rate} $(\Gamma)$ & {\bf Stability Condition}\\
			\hline
			$1+\gamma >0$ , ~$\omega >0$ (non-phantom) & There is no constraint on particle annihilation if $\gamma > 0$ but if there is particle creation then for $\gamma > 0$ we have $0 < \Gamma < \frac{3H\gamma}{1+\gamma}$\,. If $\gamma < 0$ then there should be only particle annihilation with restriction\,: $\left|\Gamma \right| > \frac{3H\left|\gamma\right|}{1+\gamma}$ & $\frac{d\ln \omega}{d\ln V}< \omega ~~\mbox{or}~~ \frac{d\ln \omega}{d\ln V}> (1+\omega)$ \\
			\hline
			$1+\gamma >0$ , ~$\omega <0$ (non-phantom) & No particle annihilation is possible for $\gamma > 0$, rather the particle creation parameter is constraint by $\frac{3H\gamma}{1+\gamma}< \Gamma < 3H$. If $\gamma < 0$ there is no constraint for particle creation while for particle annihilation the parameter is constraint as $\left|\Gamma \right| < \frac{3H\left|\gamma\right|}{1+\gamma}$ & $\omega<\frac{d\ln \omega}{d\ln V}<(1+\omega)$ \\
			\hline
			\hspace{5mm}$1+\gamma <0$ , ~$\omega >0$\newline(phantom) & No particle annihilation. The particle creation parameter is restricted as $\Gamma > \mbox{max}\left\{3H, \frac{3H\gamma}{1+\gamma}\right\}$ & $\frac{d\ln \omega}{d\ln V}< \omega ~~\mbox{or}~~ \frac{d\ln \omega}{d\ln V}> (1+\omega)$ \\
			\hline
			\hspace{5mm}$1+\gamma <0$ , ~$\omega <0$\newline(phantom) & There is no constraint if there is particle annihilation. For particle creation the parameter is constraint as $3H < \Gamma < \frac{3H\gamma}{1+\gamma}$ & $\omega<\frac{d\ln \omega}{d\ln V}<(1+\omega)$ \\
			\hline
			
		\end{tabular}
	\end{table}
\end{center}

\vspace{-0.8cm}
The present work is an attempt to make thermodynamical analysis of the physical system describing the cosmic evolution having cosmic fluid in the form of perfect fluid with constant equation of state parameter in non-equilibrium thermodynamics with particle creation mechanism. It has been shown (in ref.\,\cite{Chakraborty:2014oya,Chakraborty:2014fia,Chakraborty:2014ora}) that particle creation mechanism has successfully described the entire cosmic evolution, so we have analyzed the stability of the system in the present work from the thermodynamical point of view and the stability conditions are presented in tabular form (see table \ref{tab:1}). From the table it has been seen that cosmic fluid may be normal $\left(\gamma > -\frac{1}{3}\right)$ or exotic $\left(\gamma < -\frac{1}{3}\right)$ in nature (even phantom fluid $(\gamma < -1)$) and there is restriction on the variation of the effective equation of state parameter as well as on the particle creation rate for the stability of the system. Also except the 3rd case both annihilation and creation of particles are possible (constrainted or unconstrainted) while for the third choice destruction of particles is not permissible. In fact, the stability conditions restrict the evolution of the effective equation of state parameter with the evolution of the Universe. Further, it is found that if the cosmic fluid is vacuum energy or $\Gamma = 3H$ then the effective fluid equation of state parameter is at the phantom barrier and stability analysis will not be possible. Thus from the point of view of thermal stability, the cosmic fluid may be of any nature (normal or exotic but not vacuum energy) to describe the cosmic evolution.

\section{Acknowledgement}

The author PB acknowledges DST-INSPIRE (File no: IF160086) and SH acknowledges UGC-JRF for awarding Research fellowship. The author SC is thankful to the Inter-University Centre for Astronomy and Astrophysics\,(IUCAA), Pune, India for research facilities at Library. SC also acknowledges the UGC-DRS Programme in the Department of Mathematics, Jadavpur University.\\\\


\end{document}